\documentclass[letterpaper, 10 pt, conference]{ieeeconf}  %

\IEEEoverridecommandlockouts                              

\usepackage{graphicx} 
\usepackage{amsmath} 
\usepackage{amssymb}  
\usepackage[usenames, dvipsnames]{color}

\usepackage{lipsum}
\usepackage{color}
\usepackage{cite}

\usepackage{diagbox} 
\usepackage{tabu}
\usepackage{balance}  

\usepackage[ruled,linesnumbered]{algorithm2e}
\usepackage{algpseudocode}
\usepackage{epsfig}

\usepackage{booktabs,tabulary,lipsum}
\usepackage{dcolumn,tipa}
\newcolumntype{d}{D{.}{.}{6.5}}
\usepackage{siunitx}
\usepackage[usestackEOL]{stackengine}
\usepackage{multirow}

\title{
Thoracic Cartilage Ultrasound-CT \\Registration using Dense Skeleton Graph
}

\author{Zhongliang Jiang*$^{1}$, Chenyang Li*$^{1}$, Xuesong Li$^{1}$, and Nassir Navab$^{1,2}$, \textit{Fellow, IEEE} 
\thanks{$^{*}$ Authors are with equal contributions.}
\thanks{$^{1}$Z. Jiang, C. Li, X. Li, and N. Navab are with the Chair for Computer Aided Medical Procedures and Augmented Reality, Technical University of Munich, Germany. {\tt\footnotesize{(zl.jiang@tum.de)}}
        }%
\thanks{$^{2}$N. Navab is also with the Laboratory for Computational Sensing and Robotics, Johns Hopkins University, Baltimore, MD, USA.}
}

\begin{document}

\maketitle


\begin{abstract}
Autonomous ultrasound (US) imaging has gained increased interest recently, and it has been seen as a potential solution to overcome the limitations of free-hand US examinations, such as inter-operator variations. However, it is still challenging to accurately map planned paths from a generic atlas to individual patients, particularly for thoracic applications with high acoustic-impedance bone structures under the skin. To address this challenge, a graph-based non-rigid registration is proposed to enable transferring planned paths from the atlas to the current setup by explicitly considering subcutaneous bone surface features instead of the skin surface. To this end, the sternum and cartilage branches are segmented using a template matching to assist coarse alignment of US and CT point clouds. Afterward, a directed graph is generated based on the CT template. Then, the self-organizing map using geographical distance is successively performed twice to extract the optimal graph representations for CT and US point clouds, individually. To evaluate the proposed approach, five cartilage point clouds from distinct patients are employed. The results demonstrate that the proposed graph-based registration can effectively map trajectories from CT to the current setup for displaying US views through limited intercostal space. The non-rigid registration results in terms of Hausdorff distance (Mean$\pm$SD) is $9.48\pm0.27~mm$ and the path transferring error in terms of Euclidean distance is $2.21\pm1.11~mm$. 
\end{abstract}



\bstctlcite{IEEEexample:BSTcontrol}
\section{Introduction}

\par
Medical ultrasound (US) imaging has been widely used in a variety of clinical applications due to its radiation-free, noninvasive, and real-time characteristics. Furthermore, the low cost and availability of US make it become the primary tool for routine screening programs for internal lesions, particularly in the preliminary healthcare industry. However, traditional free-hand US examination is highly user-dependent. To alleviate inter- and intra-variations among operators, the robotic US system (RUSS) has attracted increasing attention in recent years. Owing to the high positioning and repeating accuracy, robotic techniques are employed to quantitatively control the imaging acquisition parameters~\cite{gilbertson2015force,jiang2020automaticTIE}. 

\begin{figure}[ht!]
\centering
\includegraphics[width=0.48\textwidth]{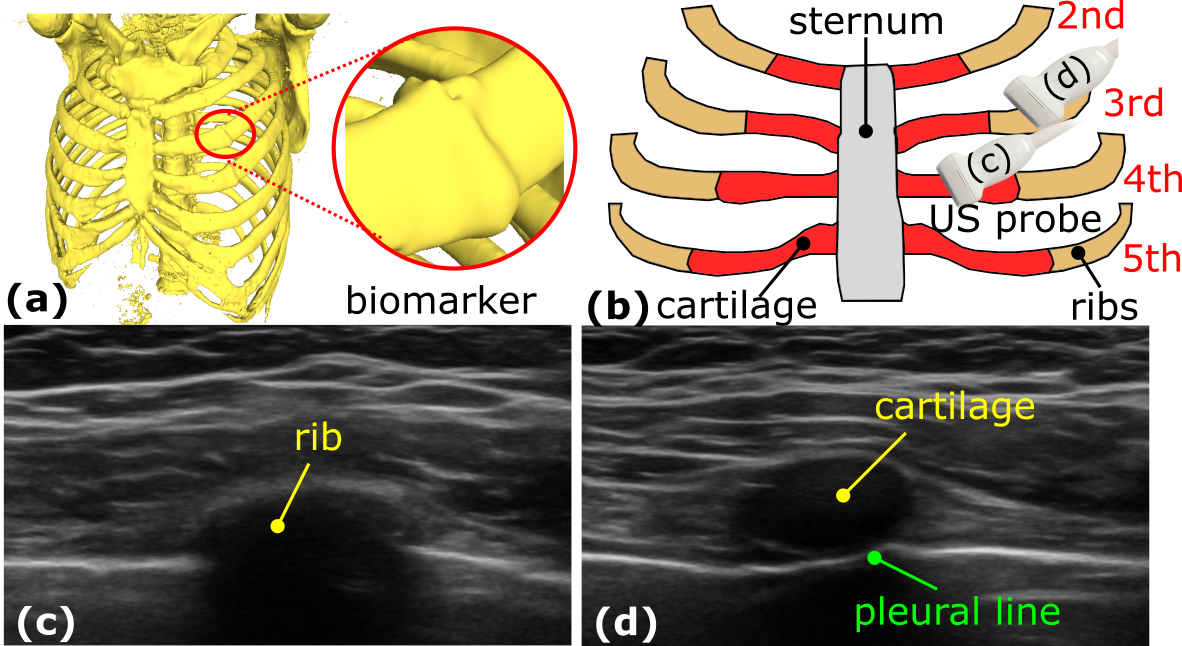}
\caption{Illustration of the biomarker of thoracic ribs on both CT and US images. (a) the connection part (biomarker) between ribs and cartilage on CT image; (b) three types of the rib cage bone: sternum, rib and costal cartilage; (c) and (d) are representative B-mode images of cartilage and rib, respectively. Due to the low acoustic impedance of cartilage, continuous subcutaneous pleural lines are shown beneath cartilage in US images.}
\label{Fig_issue_stament}
\end{figure}

\par
To determine the scanning path, Huang~\emph{et al.} extract the object of interest from RGB images and then compute a multiple-line scanning trajectory to cover the target anatomy fully~\cite{huang2018robotic}. To improve the precision of the 3D scanning path, Tan~\emph{et al.} fused the extracted surface point clouds from multiple camera views~\cite{tan2022fully}. However, these methods only consider the object's surface, whereas human subcutaneous tissues such as bone can cast shadows that obscure the anatomies' visibility. To address this issue, Sutedjo~\emph{et al.} computed the tilted poses to reduce the shadowed areas based on the scanned US volumetric data~\cite{sutedjo2022acoustic}. Considering the real applications such as visualizing the liver or heart covered by ribs, G{\"o}bl~\emph{et al.} optimized the scan poses through rib gaps to achieve the necessary coverage of target tissues on tomographic images~\cite{gobl2017acoustic}. 

\par
To transfer the planned trajectory on pre-operative images to the current setup, Hennersperger~\emph{et al.} proposed a skin surface-based registration approach to align the template CT data to the surface point cloud extracted from a RGB-D camera~\cite{hennersperger2016towards}. To further consider articulated motions around joints, Jiang~\emph{et al.} mapped the preplanned path from MRI to the current setup using non-rigid registration~\cite{jiang2022towards}. Both of these approaches are proven to be robust in their clinical scenarios (aorta and limb artery tree). However, the effectiveness of their method on clinical applications requiring intercostal views is limited without explicitly considering subcutaneous bone structures in the registration. Such views are often needed to visualize vital organs covered by ribs (such as the heart, liver, and lungs)~\cite{gordon2022b} and for guiding the intercostal puncture biopsy~\cite{he2023preoperative, rouviere2006mr}.

\par
Due to the invariance of bone structures on images, it is considered as a robust biometric feature that can be used to properly transfer a preplanned scanning path. After extracting the 3D bone surfaces from source and target images (e.g., CT/US and US/US), the classical iterative closest point (ICP)~\cite{besl1992method} can be used to optimize the transformation matrix. Considering the variation of objects, the non-rigid ICP~\cite{amberg2007optimal} and the Coherent Point Drift (CPD) algorithm~\cite{myronenko2010point} were proposed for non-rigid point set registration. Since the point clouds may only be partially observed, Zhang~\emph{et al.} proposed the partially reliable normal vectors to cast the registration problem as a maximum likelihood estimation problem~\cite{zhang2021reliable}. The method has been proven effectively work on a femur head bone; however, the performance may decrease if the object does not have significant geometry changes. To reduce the strict need for the pre-scanned CT or MRI from the same patient, deformable registration~\cite{min2020robust} can be applied to enable registration between patient-specific US data and generic templates. To monitor the growth of plants, Chebrolu~\emph{et al.} explicitly segmented leaves and stem, and then presented a spatio-temporal non-rigid registration approach based on the key points defined on both leaves and stem~\cite{chebrolu2020spatio, magistri2020segmentation}. We take inspiration from this work to design a novel graph-based non-rigid registration between generic CT images and patient-specific US images for thoracic applications with limited acoustic windows. Unlike the cases in~\cite{chebrolu2020spatio, magistri2020segmentation}, where the objects (leaves and stern) can be fully observed and segmented, the rib cage cannot be fully visualized only from the front side using US imaging. To ensure the same region of interest (ROI) can be selected on both CT and US images, and even different patients, the differences between cartilage bone and ribs are used as biomedical markers (see Fig.~\ref{Fig_issue_stament}).

\par
In this study, we present a non-rigid graph-based registration by explicitly considering subcutaneous thoracic cartilage features extracted from both a tomographic template and US images of patients. Compared to the published studies using skin surface for mapping scanning path from CT/MR to the current setup~\cite{hennersperger2016towards, jiang2022towards}, the subcutaneous bone structures with high acoustic impedance can better represent US imaging scenarios in thoracic applications by capturing the intercostal space. Therefore, we can expect the graph-based registration method to properly transfer planned paths to the current setup. The main contributions are summarized as follows:
\begin{itemize}
  \item A novel non-rigid registration method is developed based on a skeleton graph that enables robust and accurate transferring of scanning paths from tomographic templates to the current setup captured using US images. The method is based on the use of invariant subcutaneous bone surfaces and is particularly useful in thoracic applications where acoustic windows (intercostal space) are limited.
  
  \item A directed skeleton graph template is generated and updated successively twice using the self-organizing map algorithm to accurately characterize the topologies of CT and US cartilage point clouds, respectively. To avoid the potential interference between neighboring cartilage branches, the geographical distance is used instead of Euclidean distance. 
  
  \item A weighted transformation formula is presented for mapping individual CT points to US space while simultaneously preserving local accuracy and anatomy continuity.
  
\end{itemize}
Finally, the proposed graph-based non-rigid registration is validated on the US data recorded from a volunteer and five CT images from different patients. The results demonstrate that the proposed method outperforms the classical ICP, non-rigid ICP, and CPD algorithms in terms of both point cloud registration in terms of Hausdorff distance (Mean$\pm$SD: $9.48\pm0.27~mm$ vs. $17.35\pm3.54~mm$, $14.58\pm3.36~mm$, $14.72\pm3.27~mm$) and path transferring in terms of Euclidean distance ($2.21\pm1.11~mm$ vs. $4.78\pm2.83$, $4.28\pm2.37~mm$, $3.95\pm2.33~mm$). The source code will be updated and publicly accessed on this webpage\footnote{ https://github.com/marslicy/Cartilage-graph-based-US-CT-Registration}.

\begin{figure*}[ht!]
\centering
\includegraphics[width=0.98\textwidth]{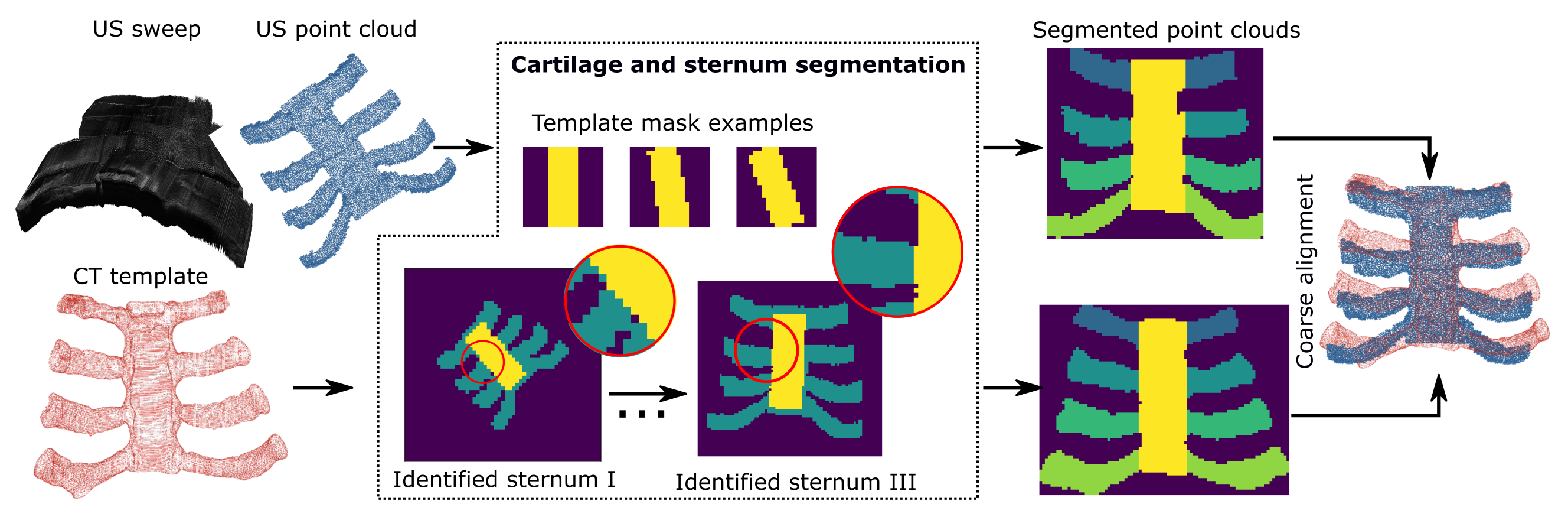}
\caption{The proposed pipeline of coarse alignment.}
\label{Fig_cosse_alignment}
\end{figure*}

\section{Preliminaries}
\subsection{Robotic US System}
\par
The RUSS consists of a collaborative robotic arm (LBR iiwa 7 R800, KUKA GmbH, Germany) and an ACUSON Juniper US System (Siemens Healthineers, Germany). A linear probe (12L3, Siemens, Germany) is attached to the robot flange. The robot is controlled via a self-developed robotic operation system (ROS) interface\footnote{https://github.com/IFL-CAMP/iiwa\_stack}. The robot status is updated at $100~Hz$. The US images are captured by a frame grabber (Epiphan video, Canada) using OpenCV in $30~fps$. To properly visualize the bone structure in B-mode images, a default setting provided by the manufacturer is used in this study: MI: $1.13$, TIS: $0.2$  TIB: $0.2$ DB: 60 dB. Since the ribs of interest are shallow, the imaging depth was set to $35~mm$.

\subsection{Robotic US Calibration}
\par
In order to stack 2D B-mode images into 3D space, we attached a US probe to the robotic end-effector. Based on the built-in kinematic model, the live tracking stream of the tool center point (TCP) can be extracted. Since our setup is similar to~\cite{jiang2021autonomous}, the same method is used here to compute the transform matrix mapping the pixel from B-model images to the robotic base frame.

\par
To preserve the anatomy's geometry accurately, we also need to synchronize the B-mode images and the robotic tracking data. Thereby, a temporal calibration is necessary, particularly when a multi-line trajectory is used. Inspired by the process presented in~\cite{salehi2017precise}, we periodically move the robot upward and downward on a gel phantom containing a 3D-printed landmark. Based on the recorded B-mode images, the landmark's movement trajectory in terms of pixel location can be extracted. In addition, the robot movement trajectory in 3D space can be directly obtained based on the forward kinematic model. By maximally matching the change tendency, the temporal shift between the robotic tracking data and the US images can be synchronized. In our setup, the computed temporal difference was $268~ms$.

\subsection{Cartilage Point Cloud Generation} \label{sec:II_Point_Cloud_Generation}
\par
Regarding point set registration, particularly for the non-rigid ones, it is vital to select an identical ROI, such as intact organs from different patients or partial observations of the same objects~\cite{min2020robust}. For clinical tasks requiring intercostal US views, benefiting from the biomarker of cartilage bone on both CT and US images (see Fig.~\ref{Fig_issue_stament}), we can extract the intact cartilage bones from various patients. Due to the invariant characteristics of bone, the scan path generated on a template CT can be projected to the current setup using US images based on the proposed graph-based cartilage bone registration approach. 


\subsubsection{Point Clouds Generation from CT Template}
\par
In this study, five CT chest scans were selected from a publicly accessible dataset\footnote{https://github.com/M3DV/RibSeg}. We manually extracted the cartilage structures, including the sternum, from CT data. The detailed procedures have been described as follows: 1) an intensity threshold-based segmentation is performed in an open-source software 3D Slicer\footnote{https://www.slicer.org/} to extract rib cages [see Fig.~\ref{Fig_issue_stament}~(a)]. 2) Then, we manually extracted the ROIs (the cartilage bones of the $2$-nd, $3$-rd, $4$-th, and $5$-th ribs) from different CTs by locating the biomarker on each rib branch in Meshlab\footnote{https://www.meshlab.net/}. 3) finally, CT point clouds  $\textbf{P}_{ct}$ could be generated by applying Poisson disc sampling. A representative example of $\textbf{P}_{ct}$ can be found at the bottom left of Fig.~\ref{Fig_cosse_alignment}.


\subsubsection{Point Clouds Generation from US Sweep}
\par
To generate patient-specific cartilage point clouds, we first manually carried out a multiple-line US scan to cover the front chest. During this process, the probe pose is represented by the robotic tracking data. Based on the one-time spatial and temporal calibration results, the recorded 2D B-mode images can be properly stacked in 3D space (see the top left of Fig.~\ref{Fig_cosse_alignment}). To further extract the cartilage part, the upper surface of the cartilage bone was manually annotated on individual B-mode images using ImFusionSuite (ImFusion AG, Germany). The cartilage bone is easy to be identified due to the visible pleural line. Then, the US cartilage point cloud (including sternum) $\textbf{P}_{us}$ can be generated by matching the proper tracking information to individual binary masks. A representative example of $\textbf{P}_{us}$ is depicted in Fig.~\ref{Fig_cosse_alignment}.

\section{Graph-based Non-Rigid Registration}
\par
After extracting the cartilage point cloud from a template CT and patient-specific US images, individually, this section presents a novel graph-based non-rigid registration approach. The registration is accomplished by successively applying the coarse alignment based on the segmented cartilage sections and sternum part, and the fine optimization based on a modified self-organizing map (SOM) algorithm~\cite{kohonen1990self}. Based on the nodes' correspondences between two computed SOM graphs, the planned path on a generic template can be properly mapped to the current setup for a specific patient. 

\subsection{US-CT Point Clouds Coarse Alignment}\label{coarse_registration}
\par
In our application, the cartilage point clouds of CT $\textbf{P}_{ct}$ and US $\textbf{P}_{us}$ are recorded from different patients. To alleviate the effect caused by patients' variations, an autonomous coarse alignment procedure is carried out based on prior anatomical knowledge, such as sternum shape. Similar to the requirement for explicit segmentation of leaves and stern of plants in~\cite{chebrolu2020spatio}, this study first segments cartilage bones and sternum from the point clouds. Nevertheless, due to the lack of comprehensive data on human thoracic images, this study uses a template matching approach rather than learning-based methods to extract the sternum. The detailed processes are depicted in Fig.~\ref{Fig_cosse_alignment}.

\begin{figure*}[ht!]
\centering
\includegraphics[width=0.98\textwidth]{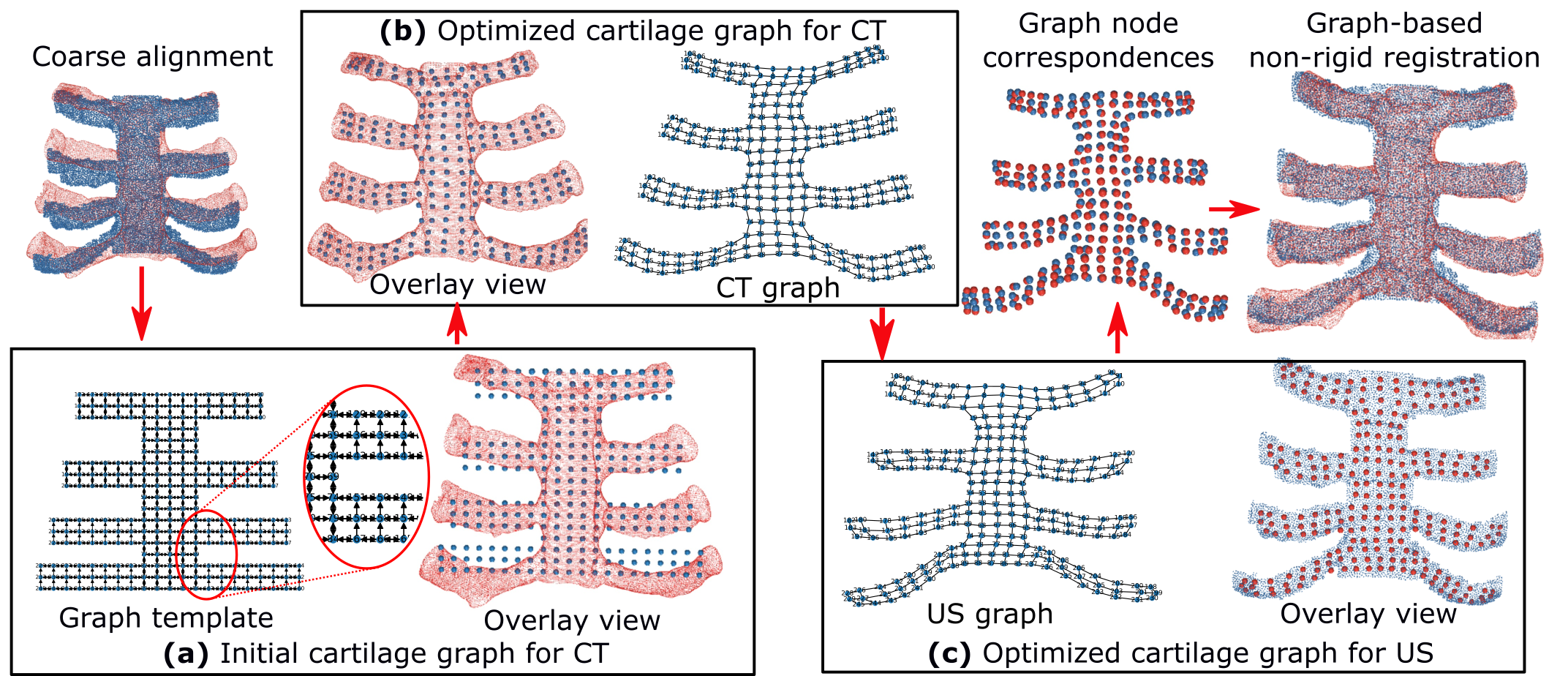}
\caption{Illustration of the proposed pipeline of the graph-based non-rigid registration.}
\label{Fig_fine_alignment}
\end{figure*}

\subsubsection{Sternum Detection and Segmentation}
\par
Considering the geometry of the cartilage bone cage, the main structure will be preserved from the front view. This characteristic could simplify the problem from 3D to 2D spaces. To obtain the front view of CT or US cartilage point clouds recorded in different coordinate systems, the principal components analysis (PCA) algorithm~\cite{abdi2010principal} is carried out to compute the three main directions of input point clouds. Then, the plane being normal to PCA's third principal direction is employed as the feature plane to segment the cartilage and sternum. A typical view of the feature plane can be seen in Fig.~\ref{Fig_cosse_alignment}. Since the sternum's projection on the feature plane (2D) will maintain a rectangle-like shape, a number of rectangle binary masks with varying orientations are generated to allow paralleling computation. The rectangle template's shape (width$\times$height) is determined based on the average measurements of sternums on CT templates. It is $30\times90~mm$ in this study. To map real-world dimension into pixels, a relatively large scale is set as $1~pixel = 5~mm$ at beginning to guarantee the time efficiency. Then, a square mask $M$ ($m\times m$) with the edge length of $m=$max(width, height) is created. Then different masks can be generated by rotating the sternum rectangle mask around the center of $M$. Three examples of the template masks (rotation: 0, $10^{\circ}$, and $20^{\circ}$ in counterclockwise direction) are visualized in Fig.~\ref{Fig_cosse_alignment}. After equal padding the feature plane, each template mask $M$ is used as a kernel for the convolution operation as follows:
\begin{equation}\label{eq_mask_template}
F_k = \max_{x \leq x_{max}; y\leq y_{max}} \sum_{i=1}^{m}\sum_{j=1}^{m}\left[M(i,j)\cdot I_b(x+i, y+j) \right]
\end{equation}
where $F_k$ is the matching value for $k$-th template, $I_b$ is the binary represent of the 2D projection on the feature plane, $x$ and $y$ is the position of the kernels' top left corner on the feature plane, $x_{max}$ and $y_{max}$ is the width and height of the unpadded feature plane. For the initial identification, $m=18=\frac{90~mm}{5~mm}$, $1\leq k \leq 18=\frac{180^{\circ}}{10{\circ}}$. By finding the optimal template achieving the highest $F_k$, we can roughly locate the pixel position of the sternum as the ``identified sternum I" in Fig.~\ref{Fig_cosse_alignment}.

\par
To further improve the boundary accuracy of segmented objects, ``identified sternum I" is used as the initial alignment. Then, the aforementioned procedures are repeated two more times with smaller parameters: scale $1~pixel = 3~mm$ and $1~pixel = 2~mm$; rotation searching ranges: $[-15^{\circ}, 15^{\circ}]$ and $[-7^{\circ}, 7^{\circ}]$; rotation step for generating mask templates: $2^{\circ}$ and $1^{\circ}$. A representative result after ruining the searching process three times is shown as ``identified sternum III" in Fig.~\ref{Fig_cosse_alignment}.

\subsubsection{Cartilage Structure Segmentation and Classification}
\par
Based on the extracted sternums, we further extract cartilage structure from the point cloud to facilitate a good coarse alignment. To this end, we first extract the centerline of the rectangle sternum mask (see ``identified sternum III" in Fig.~\ref{Fig_cosse_alignment}) parallel to the long edge. To identify the boundary between cartilage and sternum, the centerline is moved bidirectionally along the short edge direction until the number of overlapped pixels between the moving line and the rib cage ($I_b$) is dramatically decreased. 

\par
After locating the sternum-cartilage boundaries on both sides, the classic K-Nearest-Neighbors (KNN) algorithm is used to cluster the pixels of $I_b$ near the boundaries in terms of Euclidean distance, individually. The number of the cluster is determined to be four because there are four levels of cartilage involved in this work. Based on the computed eight center points of KNN, the flood fill algorithm~\cite{agkland1981edge} was repeatedly used eight times to segment the isolated cartilage branches on both sides. The flood fill algorithm can extract the area connected to a given node in 2D images. To avoid inadvertently including points from the sternum, the cartilage and sternum regions were assigned distinct attributes. The segmented CT and US point clouds can be seen in Fig~\ref{Fig_cosse_alignment}. Due to the prior knowledge that the curvature of the $5$-th cartilage is much larger than the one of $2$-nd cartilage, we can further identify the level of cartilage structures. 

\par
In the 2D view on the feature plane, we first accurately overlay the extracted sternums from CT and US. Then, the information of cartilage level is used to properly align the orientation of two point clouds. After assigning depth values in the normal direction of the feature plane back to the each point in 2D view, the initial alignment of US and CT point clouds in 3D space can be achieved. A representative result can be seen in Fig~\ref{Fig_cosse_alignment}.




\subsection{Graph-based Node Correspondence Optimization}
The pipeline of the graph-based non-rigid registration is described in Fig.~\ref{Fig_fine_alignment}. To optimize the paired point correspondences, the modified SOM using geographical distance is applied twice. Then, the local transformation of individual points among $\textbf{P}_{ct}$ is computed based on neighboring paired correspondences to map $\textbf{P}_{ct}$ to $\textbf{P}_{us}$. 

\par
Based on the extracted CT point clouds, a generic directed template graph $G_{tmp}$ is created as Fig.~\ref{Fig_fine_alignment}. The number of nodes for each parts of rib cage are determined based on the physical measurements from the CT template. In total, $245$ numbered nodes are used in this study. To accurately capture the topological structure, the SOM algorithm~\cite{kohonen1990self} is used to update the position of individual nodes. The SOM is an unsupervised machine learning method trained using competitive learning. In order to update the graph, the weight vector $\textbf{W}_s$ for each node is computed between the nodes and a random sample of the input point cloud in terms of Euclidean distance. The node with the smallest weight is called the best matching unit (BMU). To avoid misassignment of the nodes to distinct parts, the geographical distance of directed
$G_{tmp}$ is used to determine the update rate for each node. Intuitively, due to the relatively large geographical distance between the nodes from different cartilage branches, they would not affect each other. $\textbf{W}_s$ of each node is updated as follows:
\begin{equation}\label{eq_som}
\textbf{W}_{s}(i+1)=\textbf{W}_{s}(i)+\theta_{(BMU, i)}\cdot l_r\cdot \left[\textbf{P}(k)-\textbf{W}_{s}(i)\right]
\end{equation}
where $i$ is the current iteration, $\theta_{(BMU, i)}$ is the updated restriction function computed on the geographical distance between the winner node and other nodes, $l_r$ is the learning rate, and $\textbf{P}(k)$ is the $k$-th point in the point cloud. 

\par
The optimized graph for CT $G_{ct}$ is shown in Fig.~\ref{Fig_fine_alignment}. The overlaid $G_{ct}$ and $\textbf{P}_{ct}$ image demonstrates the optimized $G_{ct}$ can properly capture the characteristic of $\textbf{P}_{ct}$. Since $\textbf{P}_{ct}$ and $\textbf{P}_{us}$ have already been coarsely aligned, the optimized $G_{ct}$ is used as the initial graph for computing the optimal graph for US $G_{us}$. The computed $G_{us}$ can be seen in Fig.~\ref{Fig_fine_alignment}. Based on the numbered $G_{ct}$ and $G_{ct}$, the paired node correspondences can be achieved. It is also worthy noting that the connections between two nodes in $G_{tmp}$ are bidirectional, besides the ones belonging to the $4$-th and $5$-th cartilage branches. In these two branches, the vertical connections are in single direction (see the zoom view in Fig.~\ref{Fig_fine_alignment}). This feature is used to adapt the relatively large natural curve of cartilage. Intuitively, the nodes in the upper side can only be pulled in a downward direction during SOM updating, while the inverse cases are forbidden. After three epochs at beginning with a learning rate of $l_r=0.02$, the nodes arranged for cartilages are pulled to roughly correct locations. The directed graph is converted to indirect, and the learning rate is set to $l_r=0.1$ for the rest of the SOM process to accurately capture the geometry of CT or US data.

\subsection{Graph-based Non-Rigid Registration}
\par
Based on the optimized node correspondences, the local transformation of individual nodes of graph $G_{ct}$ can be computed based on the closest $N_{reg}$ nodes in terms of geographical distance. The local transformation matrix $^{us}_{ct}\textbf{T}$ mapping $^{ct}\textbf{P}_{g}$ to $^{us}\textbf{P}_{g}$ can be optimized by minimizing Eq.~(\ref{eq_rotatopn_optimization}).
\begin{equation} \label{eq_rotatopn_optimization}
\min_{^{ct}_{us}\textbf{T}} \frac{1}{N_{reg}}\sum_{i=1}^{N_{reg}}{||^{us}_{ct}\textbf{T}~^{ct}\textbf{P}_{g} - {^{us}\textbf{P}_{g}}||^2}
\end{equation}
where $^{ct}\textbf{P}_{g}$ and $^{us}\textbf{P}_{g}$ are the positions of paired nodes from $G_{ct}$ and $G_{us}$, respectively. A large $N_{reg}$ will reduce the non-rigid probability. An extreme case is that if we use all nodes to calculate the transformation matrix, the method will be degraded to a rigid registration. Here $N_{reg}=3$ to obtain high local accuracy. In order to preserve the object continuity, a weighted transformation method is used for transferring the CT point cloud to US space as follows: 
\begin{align}\label{registration_eq}
    ^{ct}\mathbf{P}^{\prime}=\sum_{i=1}^N \frac{d_i}{\sum_{j=1}^N d_j}\left(^{us}_{ct}\textbf{T}_i [^{ct}\mathbf{P}; 1]^{T}\right)
\end{align}
where $d_{i,~\text{or}~j}$ is the Euclidean distance between individual point among $\textbf{P}_{ct}$ and the closet $N$ nodes in $G_{ct}$, $^{ct}\mathbf{P}^{\prime}$ is the transformed CT point in US space.

\subsection{Path Transferring from CT to US}

\par
Considering the task of autonomous intercostal path transferring for thoracic applications, a similar process is carried out to map the planned trajectory on the CT template to the US setup. Because the path is located in the intercostal spaces, relatively large sphere regions are used to include enough points for mapping trajectory from CT to US. In this study, we empirically set the radius of sphere regions to $20~mm$ based on the performance. The sphere center is determined to be individual waypoints on the planned trajectory on CT. Based on the points inside the region of interest from $\textbf{P}_{ct}$ and their paired points on the transferred $^{ct}\textbf{P}^{\prime}$ [computed in Eq.~(\ref{registration_eq})], the local optimal rigid transformation matrix can be computed.

\begin{figure*}[ht!]
\centering
\includegraphics[width=0.85\textwidth]{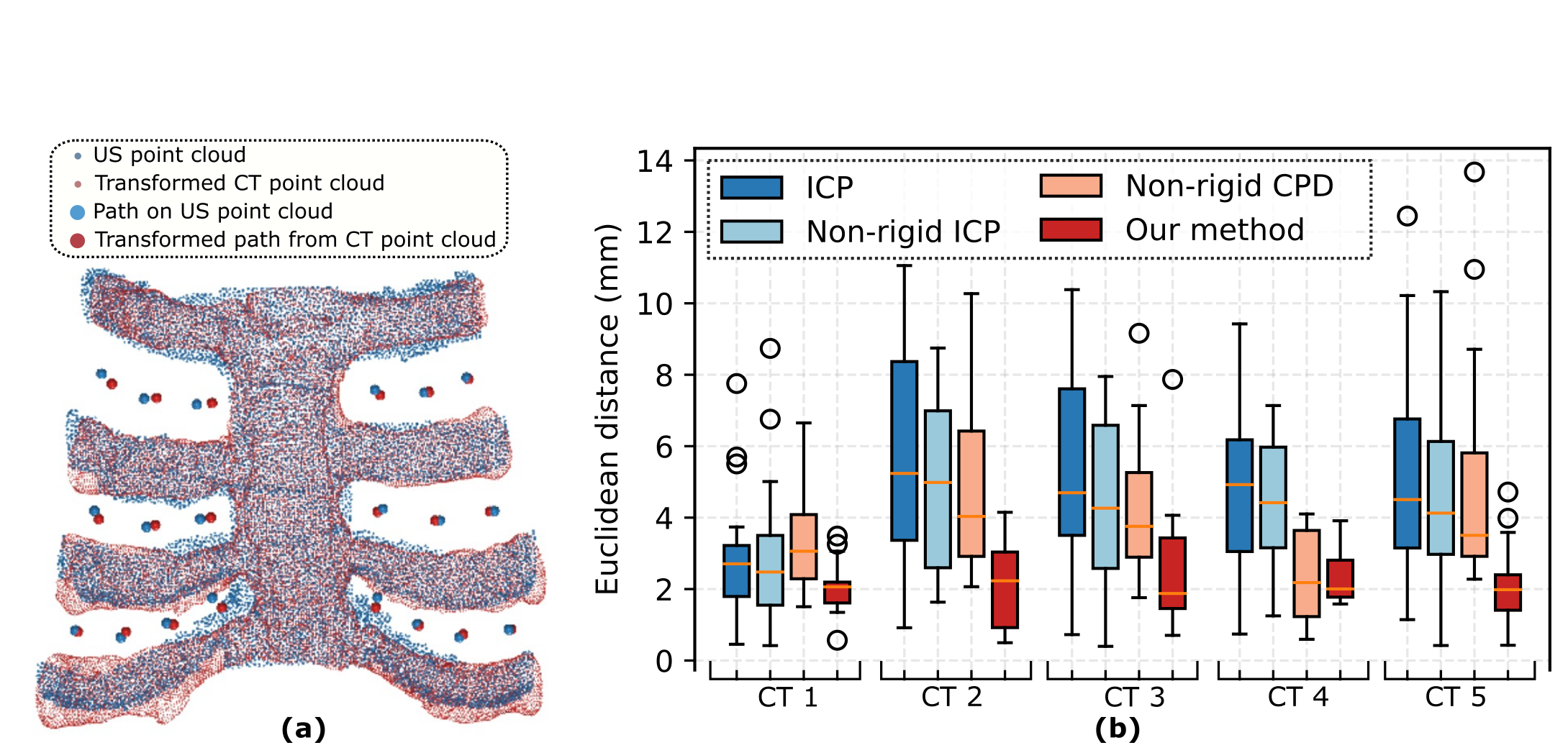}
\caption{Performance of scanning paths transferring from CT to US space on five distinct CT images. (a) Illustration of the results of $20$ waypoints distributed in all intercostal spaces mapped from CT to US. (b) The statistical results of the mapping performance in terms of position error.}
\label{Fig_path_transfer}
\end{figure*}

\section{Results}

\subsection{Performance of the Graph-based Registration}
To evaluate the performance of the proposed non-rigid registration approach, we compared our method with classic ICP algorithm~\cite{besl1992method}, non-rigid ICP~\cite{amberg2007optimal}, and CPD algorithm~\cite{myronenko2010point}. To ensure the proposed graph-based non-rigid registration method has the potential to adapt to patient-specific properties, five chest CT images from different patients are employed in this study. The manually segmented five CT point clouds are registered to a US point cloud obtained from a healthy volunteer (gender: male, age: $26$, BMI: $19.8$, height: $174~cm$) in our setup. Since there is no explicit correspondences in terms of individual points in $\textbf{P}_{ct}$ and $\textbf{P}_{us}$, Euclidean distance is not optimal here to evaluate the registration performance. Therefore, the Hausdorff distance is employed to quantify two point clouds' similarity. The results achieved by different methods are summarized in Table~\ref{table_registration_results}. The Hausdorff distance (HD) between the transformed CT point cloud $^{ct}\textbf{P}^{\prime}$ and US point cloud $^{us}\textbf{P}$ is calculated as follows: 
\begin{equation}~\label{eq_hausdorff}
HD=\max \left\{\max _{x \in ^{ct}\mathbf{P}^{\prime}} d(x, ^{us}\mathbf{P}), \max_{y \in ^{us}\mathbf{P}} d(^{ct}\mathbf{P}^{\prime}, y)\right\} 
\end{equation}

\begin{table}[ht]
 \sisetup{
 table-number-alignment = center,
 table-figures-integer = 1,
 table-figures-decimal = 4
 }
 \begin{center}
 \caption{Registration performance of Different Approaches (HD)}
 \centering
 \renewcommand\footnoterule{\kern -1ex}
 \renewcommand{\arraystretch}{1.3}
\label{table_registration_results}
\resizebox{0.48\textwidth}{!}{
\begin{tabular}{lcccccc}
\toprule
\multicolumn{1}{c}{Subjects} & CT 1 & CT 2 & CT 3 & CT 4 & \multicolumn{1}{l}{CT 5} & Mean$\pm$SD \\ \hline
ICP & 10.92 & 20.67 & 18.35 & 16.50 & 20.29 & 17.35$\pm$3.54 \\
Non-rigid ICP & 14.45 & 10.49 & 12.43 & 15.06 & 20.48 & 14.58$\pm$3.36 \\
Non-rigid CPD & 11.75 & 15.96 & 15.62 & 10.56 & 19.71 & 14.72$\pm$3.27 \\
Our Method & \textbf{9.81} & \textbf{9.53} & \textbf{9.06} & \textbf{9.69} & \textbf{9.29} & \textbf{9.48$\pm$0.27} \\ \bottomrule
\multicolumn{6}{l}{Unit: mm} & 
\end{tabular}
}
 \end{center}
 \end{table}
 
\par
It can be seen from TABLE~\ref{table_registration_results} that the proposed graph-based method achieves the best performance for all CT images in terms of HD. The best result is achieved by the proposed method on CT 3 ($9.06~mm$). Although the other methods can sometimes achieve close values, such as ICP on CT 1 ($10.92~mm$) and non-rigid ICP on CT 4 ($10.56~mm$), the results obtained by the proposed approach are much more stable among all five CT data ($9.48\pm0.27~mm$). Therefore, we consider the proposed non-rigid graph-based registration can provide robust results in adapting inter-patient variations.

\subsection{Performance of Path Transferring from CT to US}
\par
To validate whether the proposed graph-based registration can appropriately map a planned trajectory from the generic template CT to the current US setup, $20$ waypoints of the scanning path are determined in both CT and US space following the same protocol. To ensure the defined waypoints from CT and US point clouds are the ones that should be matched after registration, we manually segmented each rib cartilage to guarantee accuracy for validation. Then, KNN is used to autonomously cluster each cartilage branch equally into multiple parts. By connecting the corresponding centers on neighboring cartilage successively, the midpoints of the connected lines are located in intercostal spaces; thereby, they are further used as waypoints of the scan path. Considering the length of cartilage, three waypoints are defined in both the first and second intercostal spaces, while four waypoints are in the third intercostal space. To avoid biased results related to the path location, in total $20$ waypoints are generated on both sides of the cartilage rib.


\par
A representative result is intuitively depicted in Fig.~\ref{Fig_path_transfer}, where the red points are transferred from the CT to the US point cloud obtained from a volunteer. The transformed waypoints are located close to the ones defined directly on US point cloud. To quantitatively evaluate the performance, the Euclidean distance between the transformed waypoints from CT and the ones defined on US point cloud was computed. To demonstrate the effectiveness of the proposed method, ICP, non-rigid ICP, and CPD are also used here. Regarding the latter two non-rigid methods, the transferring of the waypoints from CT to US is in the same way as this approach by defining sphere region with a radius of $20~mm$. The results are summarized in Fig.~\ref{Fig_path_transfer}. 

\par
It can be seen from Fig.~\ref{Fig_path_transfer} that the proposed method can result in more stable and accurate transferring results compared to others. The proposed method achieves the best performance on all five CTs in terms of both maximum and mean distance errors. The ICP obtains the worst performance among its peers in most cases (four out of five CTs). We consider the only exception case to be because the size and shape of CT 1 are very close to the used US point cloud. In this case, all methods achieve the best performance on CT 1 compared to themselves. Although the results of the non-rigid ICP (Mean$\pm$SD: $4.28\pm2.37~mm$) are slightly worse than the one obtained by CPD ($3.95\pm2.33~mm$) among all five CTs, the computation efficiency of CPD is much worse (around $20~min$ per CT). In addition, the average mapping error of ICP and the proposed method are $4.78\pm2.83~mm$ and $2.21\pm1.11~mm$, respectively. Thereby, we consider the proposed approach can outperform the existing methods in terms of adapting inter-patient variations. 

\par
\section{Discussion}
The proposed skeleton graph-based non-rigid registration has been validated on the cartilage point clouds obtained from a volunteer and five CT images from different patients. The results demonstrate that the proposed method outperforms popular methods like classic ICP, non-rigid ICP, and CPD algorithms in terms of both registration accuracy and path transferring accuracy on all CTs. Due to the absence of a camera in our setup, as a side benefit, our method does not require a hand-eye calibration procedure, which could make clinical practice more convenient. However, some limitations are worth to be discussed to inspire future studies further. First, in order to obtain the US images from the transformed path from CT images, the patients should not move after the acquisition of US point clouds. Besides, since the cartilage ribs are relatively flat, the 3D point clouds do not have very informative data in the direction normal to the front view. Thereby, the accuracy of the orientation mapping cannot be guaranteed yet. To address this challenge, internal organs or biomarkers could be considered as complementary dimensional information to the proposed cartilage graph-based non-rigid registration method.



\section{Conclusion}
This work presents a skeleton graph-based non-rigid registration approach between a tomographic template and B-mode images of patients for autonomous transferring of US scan trajectory from CT to the current setup. Compared to the existing studies using skin surface point clouds~\cite{hennersperger2016towards, jiang2022towards}, subcutaneous bone surfaces are used to better characterize challenging scenes of thoracic application with limited acoustic windows; therefore achieving better registration accuracy. The proposed method has been tested on the US data recorded from a volunteer and five CT images from a public dataset. The results demonstrate that the proposed method outperforms the classical ICP, non-rigid ICP, and CPD algorithms for point cloud registration in terms of Hausdorff distance (Mean$\pm$SD: $9.48\pm0.27~mm$ vs. $17.35\pm3.54~mm$, $14.58\pm3.36~mm$, $14.72\pm3.27~mm$) and for path transferring in terms of Euclidean distance ($2.21\pm1.11~mm$ vs. $4.78\pm2.83$, $4.28\pm2.37~mm$, $3.95\pm2.33~mm$). The results show that the proposed graph-based registration algorithm can properly adapt to inter-patient variations. This characteristic can further pave the way for addressing the challenge of robustly and autonomously generating scan paths for different patients. Thereby, further facilitating extensively deploying autonomous screening systems for challenging thoracic applications requiring intercostal US images in clinical practices. 



\bibliographystyle{IEEEtran}
\balance
\bibliography{IEEEabrv,references}

\end{document}